\documentclass[aps,prl,twocolumn,groupedaddress,showpacs]{revtex4}
\usepackage{amsmath}
\usepackage{amsfonts}
\usepackage{amssymb}
\usepackage{graphicx}
\usepackage{epsfig}
\begin{document}

\title{Kinetics of phase separation in the driven lattice gas: Self-similar pattern
growth under anisotropic nonequilibrium conditions}
\author{P.I. Hurtado}
\author{J. Marro}
\author{P.L. Garrido}

\affiliation{Instituto Carlos I de F\'{\i}sica Te\'{o}rica y Computacional,\\
and Departamento de Electromagnetismo y F\'{\i}sica de la Materia,\\
Universidad de Granada, E-18071-Granada, Espa\~{n}a}

\author{E.V. Albano}

\affiliation{Instituto de Investigaciones Fisicoqu\'{\i}micas Te\'{o}ricas y
Aplicadas, \\ 
UNLP, CONICET, La Plata, Argentina}


\date{October 7, 2002}

\begin{abstract}
The driven lattice gas (DLG) evolving at low temperature helps understanding
the kinetics of pattern formation in unstable mixtures under anisotropic
conditions. We here develop a simple theoretical description of kinetics in
Monte Carlo simulations of the DLG. A Langevin continuum analog is also
studied which is shown to exhibit the same behavior. We demonstrate that
pattern growth is mainly a consequence of single-particle processes and that,
after a short transient time, in which a surface evaporation/condensation
mechanism is important, hole diffusion in the bulk becomes dominant.
Consequently, there is a unique relevant length that behaves $\ell\left(
t\right)  \sim t^{1/3}$ for macroscopic systems except at some very early
(perhaps unobservable) time. This implies sort of self-similarity, namely, the
spatial pattern looks alike, but for a (non-trivial) change of scale at
different times. We also characterize the structure factor, in which we
identify Guinier and Porod regions, and its scaling form with both time and
size. The underlying anisotropy turns out to be essential in determining the
macroscopically-emergent peculiar behavior.
\end{abstract}



\maketitle

\section{Introduction}

Many alloys such as Al-Zn, which are homogeneous at high temperature, undergo
phase separation after a sudden quench into the miscibility gap (for details,
see the reviews \cite{rev1}-\cite{rev5}, for instance). One first observes
nucleation in which small localized regions (\textit{grains}) form. This is
followed by \textquotedblleft spinodal decomposition\textquotedblright. That
is, some grains grow at the expense of smaller ones, and eventually coarsen,
while their composition evolves with time. In addition to theoretically
challenging, the details are of great practical importance. For example,
hardness and conductivities are determined by the spatial pattern finally
resulting in the alloy, and this depends on how phase separation competes with
the progress of solidification from the melt.

A complete kinetic description of these highly non-linear processes is
lacking.\cite{rev5} Nevertheless, the essential physics for some special
situations is now quite well understood. This is the case when nothing
prevents the system from reaching the equilibrium state, namely, coexistence
of two thermodynamic phases. The simplest example of this is the (standard)
lattice gas evolving from a fully disordered state to segregation into
\textit{liquid} (particle-rich phase) and \textit{gas} (particle-poor phase).
(Alternatively, using the language of the isomorphic lattice binary alloy,
\cite{ma} the segregation is into, say Al-rich and Zn-rich phases.) As first
demonstrated by means of computer simulations,\cite{prl00,rev1,rev2} this
segregation, as well as similar processes in actual mixtures exhibit time
\textit{self-similarity}. This property is better defined at sufficiently low
temperature, when the thermal correlation length is small. The system then
exhibits a \textit{single} relevant length, the size $\ell\left(  t\right)  $
of typical grains growing algebraically with time. Consequently, any of the
system properties (including the spatial pattern) look alike, except for a
change of scale, at different \textit{times}.

This interesting property is revealed, for example, by the sphericalized
structure factor $S\left(  k,t\right)  $ as observed in scattering
experiments. After a relatively short transient time, one observes that
$S\left(  k,t\right)  \sim J\left(  t\right)  \cdot F\left[  k\ \ell\left(
t\right)  \right]  .$ Taking this as a hypothesis, one may interpret $J$ and
$\ell$ as phenomenological parameters to scale along the $S$ and $k$ axes,
respectively. The hypothesis is then widely confirmed, and it follows that
$J\left(  t\right)  \sim\ell\left(  t\right)  ^{d}$ where $d$ is the system
dimension. It also follows that $F(\varkappa)=\Phi(\varkappa)\cdot\Psi\left(
\sigma\varkappa\right)  $ where $\Phi$ and $\Psi$ are universal functions. In
fact, $\Phi$ describes the diffraction by a single grain, $\Psi$ is a grain
interference function, and $\sigma$ characterizes the point in the
(density$-$temperature) phase diagram where the sample is quenched. It then
ensues that $\Psi\approx1$ except at small values of $k$, so that, for large
$\varkappa$, $F(\varkappa)$ becomes almost independent of density and 
temperature, and even the substance investigated.\cite{prl00,zahra,rev5}

The grain distribution may also be directly monitored. A detailed study of
grains in both microscopy experiments and computer simulations confirms time
scale invariance. More specifically, one observes that the relevant length
grows according to a simple power law, $\ell\left(  t\right)  \sim t^{a},$ and
one typically measures $a=1/3$ at late times. This is understood as a
consequence of diffusion of monomers that, in order to minimize surface
tension, evaporate from small grains of high curvature and condensate onto
larger ones (\textit{Ostwald ripening}). In fact, Lifshitz and Slyozov, and
Wagner independently predicted $\ell\sim t^{1/3},$\cite{LSW} which is often
observed, even outside the domain of validity of the involved
approximations.\cite{LSWbis} In some circumstances, one should expect other,
non-dominant mechanisms inducing corrections to the Lifshitz-Slyozov-Wagner
one.\cite{rev1,rev3,rev5} For instance, effective diffusion of grains
(\textit{Smoluchowski coagulation}) leads to $a=1/6,$ which may occur at early
times;\cite{toral} interfacial conduction leads to $a=1/4;$\cite{huse,puri0}
and, depending on density and viscosity, a fluid capable of hydrodynamic
interactions may exhibit crossover with time to viscous ($a=1$) and then
inertial ($a=2/3$) regimes.\cite{rev4}

Extending the above interesting picture to more realistic situations is an
open question. The assumption that the system asymptotically tends to the
coexistence of two \textit{thermodynamic} (equilibrium) phases is often
unjustified in nature. This is the case, for example, for mixtures under a
shear flow, whose study has attracted considerable attention,
e.g.\cite{critic}-\cite{corberi}. The problem is that sheared flows
asymptotically evolve towards a \textit{nonequilibrium} steady state and that
this is highly anisotropic. Studying the consequences of anisotropy in the
behavior of complex systems is in fact an important challenge (see, for
instance, \cite{mandel}-\cite{md}). Another important example is that of
binary granular mixtures under horizontal shaking. The periodic forcing 
causes in this case phase separation and highly anisotropic clustering.\cite{Mullin}

In this paper, we study in detail the kinetics of the driven lattice gas (DLG)
\cite{katz} following a deep quench. Our motivation is twofold. On one hand,
the DLG is recognized to be an excellent microscopic model for nonequilibrium
anisotropic phenomena.\cite{md} On the other, the DLG is not affected by
hydrodynamic interactions, which makes physics simpler. Our goal is timely
given that the asymptotic state of the DLG is now rather well understood, and
previous studies of kinetics altogether reveal an intriguing
situation.\cite{md}-\cite{albano} Following this pioneering effort, we here
present a new theoretical description of the essential physics during
anisotropic, nonequilibrium pattern growth. This is compared with new
extensive computer simulations. A brief and preliminary account of some of our
results was presented elsewhere.\cite{new}

\section{Model and simulation details}

The DLG consists of a $d-$dimensional, e.g., simple-cubic lattice with
configurations $\mathbf{n=}\left\{  n_{i};i=1,...,N\right\}  $. The variable
at each lattice site has two possible states, $n_{i}=1$ (\textit{particle}) or
$0$ (\textit{hole}). As for the standard lattice gas, dynamics is a stochastic
process at temperature $T$ consisting of nearest-neighbor (NN) particle/hole
exchanges. This conserves the particle density, $\rho=N^{-1}\sum_{i}n_{i},$
and depends on $\mathbf{n}.$

A distinguishing feature of the DLG is that exchanges are favored in one of
the principal lattice directions, say $\vec{x}.$ Therefore, assuming periodic
(toroidal) boundary conditions, a net current of particles is expected to set
in along $\vec{x}.$ This is accomplished in practice by defining a biased
transition rate. We shall refer here to the \textit{energy }function
$H=-4J\sum_{NN}n_{i}n_{j},$ which describes attractive interactions between
particles at NN sites, and to the transition rate (per unit time):\cite{md}
\begin{equation}
\omega(\mathbf{n}\rightarrow\mathbf{n}^{\ast})=\min\left\{  1,\operatorname{e}%
^{-\left(  \Delta H+E\delta\right)  /T}\right\}  . \label{rate}%
\end{equation}
$\mathbf{n}^{\ast}$ stands for configuration $\mathbf{n}$ after jumping of a
particle to a NN hole; $\Delta H=H\left(  \mathbf{n}^{\ast}\right)  -H\left(
\mathbf{n}\right)  $ is the \textit{energy} change brought about by the jump;
and units are such that both the coupling strength $J$ and the Boltzmann
constant are set to unity. One further defines $\delta=\left(  \mp1,0\right)
$ for NN jumps along $\pm\vec{x}$ or along any of the transverse directions,
say $\vec{y},$ respectively. Consistent with this, $\vec{E}=E\vec{x}$ may be
interpreted as a field driving particles, e.g., an electric field if one
assumes that particles are charged. (One may adopt other interpretations,
e.g., the binary alloy one.\cite{ma} Dynamics then consists of interchanges
between particles of different species, one of them favored along $\vec{x}.$)

The DLG was described as modelling surface growth, fast ionic conduction and
traffic flow, among a number of actual situations of practical
interest.\cite{md} A common feature in these situations is anisotropy, and
that steady states are out of equilibrium. Both are essential features of the
DLG induced by the rate (\ref{rate}). The only trivial case is for $E=0,$
which reduces (\ref{rate}) to the Metropolis algorithm. In this case, detailed
balance holds, and one simply has the familiar lattice gas with a unique
(equilibrium) steady state. For any, even small $E,$ qualitatively new
behavior emerges. In fact, detailed balance breaks down for $E>0$ and,
consequently, the steady state depends on $\omega(\mathbf{n}\rightarrow
\mathbf{n}^{\ast}).$ Increasing $E,$ one eventually reaches
\textit{saturation.} That is, particles cannot jump backwards, i.e., $-\vec
{x},$ which formally corresponds to an \textit{infinite field} $(E=\infty).$

The way in which the microscopic anisotropy (\ref{rate}) conveys into
macroscopic behavior is amazing.\cite{md} Consider, for simplicity, $d=2,$
$\rho=%
\frac12
$ and $E=\infty.$ The system then exhibits a critical point at $T=T_{C}%
^{\infty}\simeq1.4T_{C}\left(  E=0\right)  $, where $T_c(E=0)\approx 2.2691$, 
with novel critical
behavior.\cite{prl01,albano} Furthermore, the asymptotic, steady states below
$T_{C}^{\infty}$ do not comprise equilibrium phases. Instead, one observes a
particle \textit{current} and fully \textit{anisotropic} phases; both are
nonequilibrium features. The intensity of the current increases with $T,$ and
suddenly changes slope at $T_{C}^{\infty}$ (in fact, this property may serve
to accurately locate the critical point). The stable ordered configurations
consist of one stripe, to be interpreted as a \textit{liquid }(rich-particle)
phase of density $\rho_{L}\left(  T\right)  .$ The \textit{gas}
(poor-particle) phase of density $\rho_{G}\left(  T\right)  $ fills the
remainder of the system. Except for some microscopic roughness, the interface
is linear and rather flat, in general.\cite{interfases}

The computer evolutions reported here always begin with a completely
disordered state to simulate the system at infinite temperature. We then model
a sudden quench and the subsequent time evolution. With this aim, one proceeds
with rate (\ref{rate}) that involves the temperature $T$ at which the system
is quenched. The run is followed until one stripe is obtained (eventually, in
order to save computer time, the run was sometimes stopped before reaching the
final stationary state). The code involves a list of $\eta\left(  t\right)  $
particle-hole NN pairs from where the next move is drawn. Time is then
increased by $\Delta t=\eta\left(  t\right)  ^{-1},$ so that its unit or
\textit{MC step} involves a visit to all sites on the average.\cite{deltat}

The lattice is rectangular, $L_{\parallel}\times L_{\perp},$ with sides
ranging from 64 to 256 and, in a few cases, 512. Results concern an average
over around thousand independent runs. Due to the great computational effort
which is consequently involved, this paper describes simulations concerning a
single point of the two-dimensional DLG phase diagram. That is, most of our
evolutions are for $\rho=%
\frac12
$ and $E=\infty,$ and simulate a quench at $T=0.8T_{C}\left(  E=0\right)
\simeq0.6T_{C}^{\infty}.$ This choice is motivated by the fact that clustering
is then reasonably compact, which helps to obtain good statistics, while it
proceeds fast enough, so that one can observe full relaxation to the steady
state. In spite of this restriction, brief investigation of other points,
together with some of our observations below, led us to believe that the
validity of our results extends to a large domain around the center of the
miscibility gap; in fact, such generality of behavior has been reported for
$E=0.$\cite{rev2,prl00,toral,rev5}

\section{Growth of order\label{growth}}

The DLG exhibits different time regimes during phase separation. Though they
parallel the ones for $E=0,$ the \textit{peculiarities} induced by the
anisotropic condition are essential.

Starting from complete disorder, there is a very short initial regime in which
small grains form. The novelty is that typical grains are now fully
anisotropic, stretched along $\vec{x}.$ The grains then rapidly coarsen to
form macroscopic strings, as illustrated in fig.1. Sheared fluids (an
experimentally accessible situation that also involves both nonequilibrium
physics and anisotropy) seem to exhibit similar initial
regimes.\cite{onuki,corberi} That is, during a short time interval, they show
larger growth rate along the flow than in the other directions, which is
assumed to correspond to the initial formation of anisotropic regions.
Afterwards, sheared fluids develop string-like macroscopic domains similar to
the ones in the DLG.
\begin{figure}
\centerline{
\psfig{file=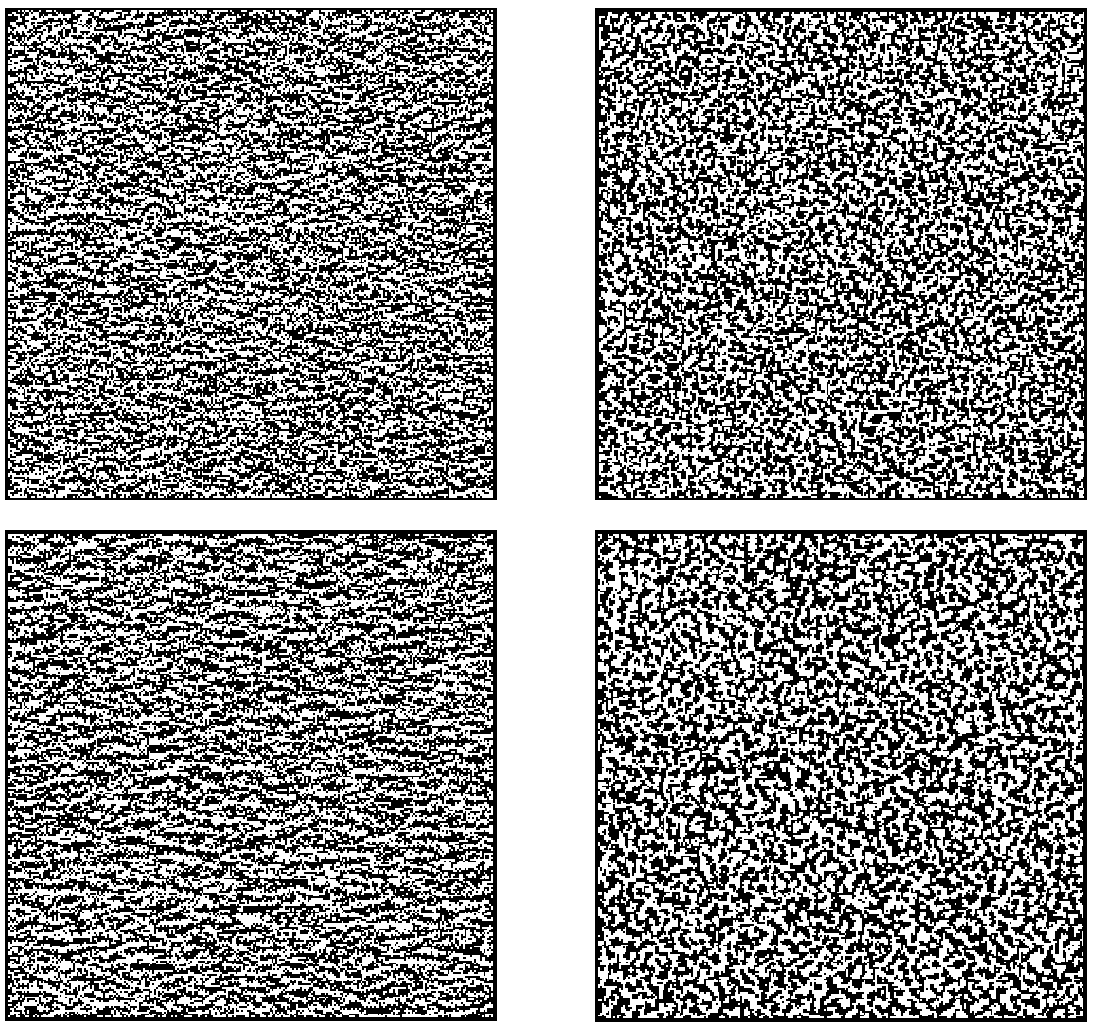,width=8.0cm,angle=0}}
\flushleft{(a)}
\flushleft{(b)}
\centerline{
\psfig{file=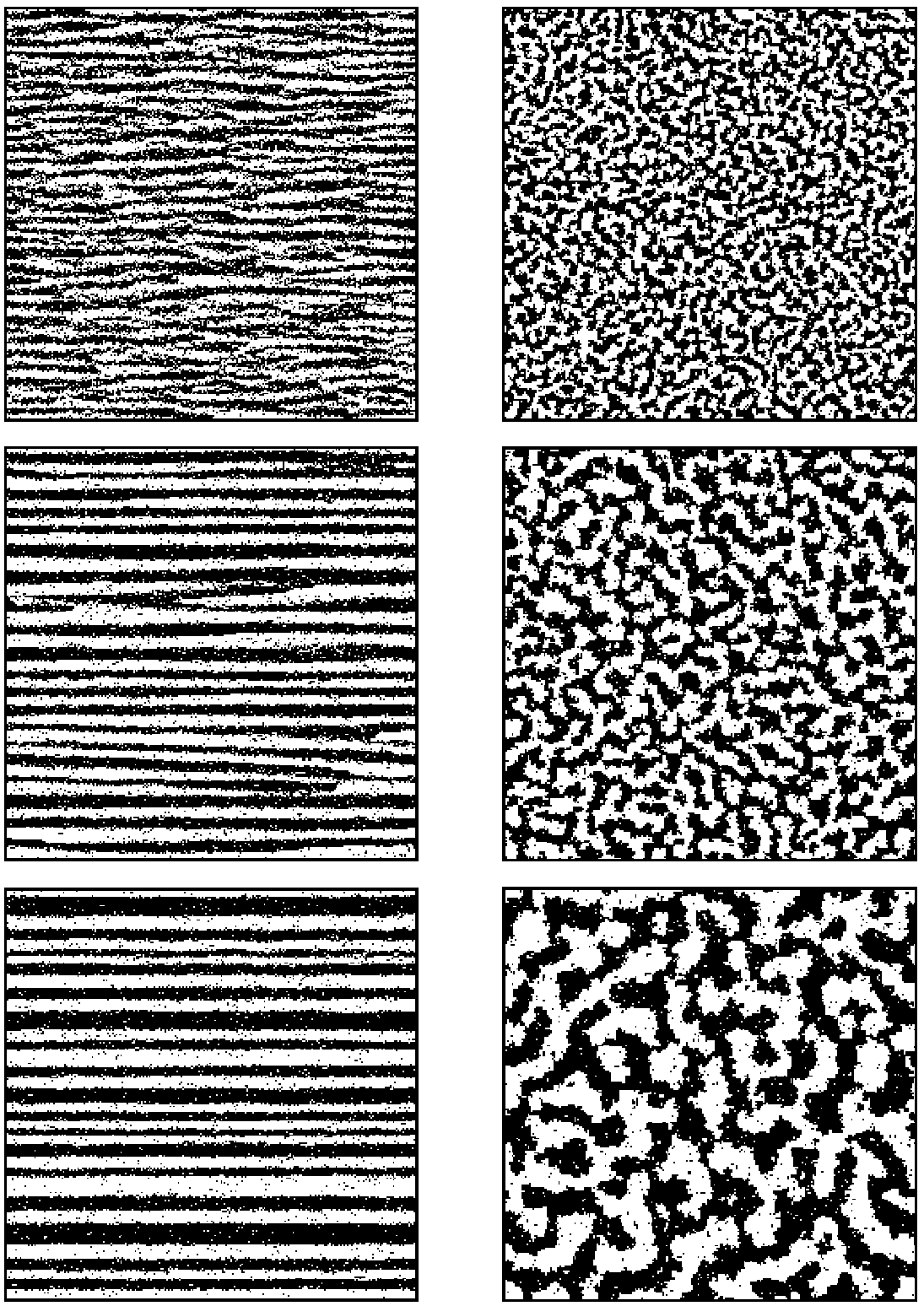,width=8.0cm,angle=0}}
\caption{\small (a) A series of MC snapshots comparing very early patterns for
the DLG (with an infinite horizontal field) and for the standard lattice gas,
i.e., zero field (LG) at the same time. This corresponds to a 256x256 lattice
at $T\simeq0.6T_{C}^{\infty}.$ The time (in MC steps) is here $t=$
4 and 10 (from top to bottom) for the DLG (left column) and for the LG
(right column). (b) The same as (a) but at late time, namely, 
$t=$ 100, 1,000 and 10,000 (from top to bottom) for the DLG (left
column) and for the LG (right column).}
\end{figure}

Fig.1 includes a comparison with the zero-field case, i.e., the standard,
isotropic lattice gas (LG). This clearly illustrates the strong anisotropy of
nucleation and early phase separation for the DLG. Close inspection of these
and similar graphs also seems to indicate relatively small but significant
differences in the degree of segregation between the two cases at a given
time. That is, at small distances, there is a more homogeneous distribution of
particles, both longitudinally and transversely, in the DLG than in the LG.
The latter shows up more segregated at the same time, which is already rather
evident by direct inspection of graphs for $1<t\leq100$ in fig.1. We believe
this reveals the different role played by surface tension as the degree of
anisotropy is varied: Typical DLG grains are rather linear except at their
longitudinal ends, where curvature may be even stronger than for the spherical
clusters in the LG at comparable times. This seems to be at the origin of a
smother transverse distribution of particles in the DLG at early times. On the
other hand, the field also tends to smooth things longitudinally.

In order to quantify the aforementioned observation, we evaluated the number 
of broken bonds in the direction of (perpendicular to) the field, $n_{\parallel}(t)$
($n_{\perp}(t)$) as a function of time during the early evolution stage. 
Then $A(t)\equiv [n_{\perp}(t)+n_{\parallel}(t)]/2N$ is
the density of broken bonds. The higher the degree of segregation at time 
$t$, the smaller is $A(t)$. For instance, we observe in a large $256\times 256$ lattice
that $A(t=10)=0.295$ and $A(t=10)=0.38$ for the LG and DLG respectively, confirming the 
above observation. On the other hand, let $B(t)\equiv [n_{\perp}(t)-n_{\parallel}(t)]/2N$. 
One would expect $B(t)\approx 0$ (up to fluctuations) only for the isotropic system.
In fact, we measured $B(t)\approx 0$ for the LG, while $B(t)$ rapidly converges to a 
nonzero value $B(t)\approx 0.05$ for the DLG at early times (again for a large 
$256\times 256$ lattice). We take this number, $B(t)=0.05$, as characterizing the 
anisotropic shape of DLG clusters at early times.

The difference of segregation between the DLG and the LG at early
times merits further study. This will need to take into account the anisotropy
of surface tension. In any case, this concerns a regime very near the initial,
melt state that only bears minor practical importance, given that it extends
extremely shortly on the macroscopic time scale. We are interested in the rest
of this paper on the subsequent evolution, to be described on the assumption
of a simple flat interface, which holds in fig. 1 for $t>100.$
\begin{figure}
\centerline{
\psfig{file=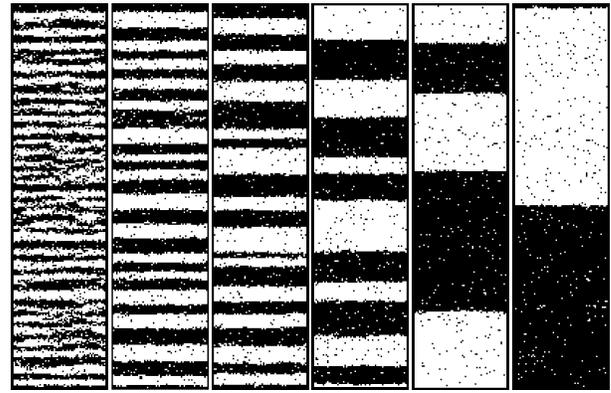,width=8.0cm,angle=0}}
\caption{\small A series of MC snapshots that illustrate (late) growth at
$T\simeq0.6T_{C}^{\infty}.$ This is for a rectangular lattice of
size $L_{\perp}\times L_{\parallel}=256\times64$ and $t=$
10$^{2}$, 10$^{4}$, 10$^{5},$ 10$^{6},$
10$^{7}$ and 1,1x10$^{8}$ MC steps,
respectively, from left to right.\medskip}
\end{figure}

The DLG strings coarsen with time until well defined, relatively-narrow
longitudinal (i.e., directed along $\vec{x})$ stripes are formed. (For
periodic boundary conditions, the case of our simulations, each stripe forms a
ring.) This results into a multi-stripe state, as illustrated in figs. 1 and
2. The ordering time in the DLG, defined 
as the time the system needs to form the stripes, scales with the system size 
in the direction of the field, $L_{\parallel}$, 
since in this case ordered clusters (stripes) 
percolate along the field direction (see below).\cite{mukamel} This is not the case 
for the equilibrium LG, where the ordering time depends exclusively on system 
intensive parameters such as temperature and density.

The multi-stripe states are not stable, however. They are only partially
segregated and, in fact, a definite tendency towards a fully segregated state
with a single stripe is generally observed in computer simulations. One may
also develop simple arguments indicating that, in general, a multi-stripe
state will monotonically evolve until forming a single stripe.\cite{25bis,md}
It is true that, in practice, the complete relaxation may take a very long
time. More specifically, a macroscopic system may take to decay into the true
stable state a long, \textit{macroscopic} time interval, namely, a time that
may show up as mathematically \textit{infinite} in some time scales. In fact,
the complete relaxation time is observed to increase with system size, as first
demonstrated in \cite{mukamel}. It should also be remarked that this property
is not a nonequilibrium feature but occurs already in the equilibrium ($E=0$)
case; see, for instance, \cite{rev2,rev4} and references therein. Slow
relaxation is a consequence of the conservation of particle density $\rho$
implied by the particle-hole exchange dynamics; this induces scale invariance,
namely, slow (power-law) evolution of correlations so that, once enough order
sets in, all but very small pattern modifications during a single MC step are
precluded. Consequently, certain individual runs sometimes block for a long
time in a state with several stripes; however, this does not correspond to the
average behavior. As illustrated by fig.2, which shows a typical evolution,
and demonstrated below by our averages corresponding to thousand evolutions,
the number of stripes monotonically decreases with time (see also
\S \ref{Lang}), and the whole relaxation can easily be observed in computer
simulations if one waits long enough.

We next attempt a theoretical description of the relaxation process. Our
interest is on the \textit{anisotropic spinodal decomposition} by which the
earliest state with many well-defined stripes decays into a single stripe.We
shall assume that relaxation is a consequence of monomer events causing
effective diffusion of liquid stripes. (Note that assuming gas stripes here
would be completely equivalent.) That is, due to single particle processes,
liquid stripes move transversely as a whole, and may collide and eventually
coalesce with one of the neighboring stripes; see the late evolution depicted
in fig.2. We notice that coalescence implies evaporation of the gas stripe
between the two involved liquid stripes. Therefore, given the particle/hole
symmetry, our assumption is in a sense equivalent to assuming that growth is
due to evaporation of stripes;\cite{mukamel} however, the view adopted here
allows for a more detailed description below.

In order to evaluate the implications of stripe effective diffusion via
monomer events, lets assume that stripes are well defined, compact and exhibit
a (linear) interface which is rather flat. This is perfectly justified at
sufficiently low temperature (the case analyzed in detail here),\cite{md} and
it might hold more generally, in a wide region including the center of the
miscibility gap but excluding the critical region. Under this assumption,
consider a stripe of mean width $\ell\left(  t\right)  $ that consists of $M$
particles whose coordinates along the transverse (vertical) direction are
$y_{j}\left(  t\right)  ;$ $j=1,\ldots,M.$ We characterize the stripe position
by its center of masses, $Y_{\text{cm}}\left(  t\right)  \equiv M^{-1}\sum
_{j}y_{j}\left(  t\right)  .$

Let us evaluate the \textit{mobility} coefficient $\mathcal{D}_{\ell}\equiv
N_{\text{me}}\langle(\Delta Y_{\text{cm}})^{2}\rangle$ which depends on the
stripe width $\ell(t).$ Here $N_{\text{me}}$ is the number of monomer events
per unit time, and $\langle(\Delta Y_{\text{cm}})^{2}\rangle$ is the mean
squared displacement of the stripe associated to one of the monomer events. We
think of two possible types of events, each giving a different contribution to
$\mathcal{D}_{\ell}:$

\noindent\ (A) Evaporation-condensation of particles and holes in the stripe surface.
Here particles (holes) at the stripe interface evaporate to the hole (particle) gas,
and condensate later at the same interface.
The evolution of the evaporated particle (hole) in the 
bulk can be seen as a one dimensional random walk with two absorbing
walls, the left and right interfaces, respectively. According to standard
random walk theory,\cite{gambler} the evaporated particle (hole) will go again 
with unit probability to one of the (two) possible interfaces, Moreover, the random 
walker will stick again to its original interface with high probability, so trapping 
a particle (hole) from the opposite interface is unlikely.
Consequently, in this case (A), $N_{\text{me}}$ is
simply the evaporation rate. That is, $N_{\text{me,A}}=\nu\sum_{j}^{\prime
}\exp(-2T^{-1}\Delta_{j}),$ where $\nu$ is the \textit{a priori} frequency,
the sum is over the surface particles and $\Delta_{j}$ is the number of
resulting broken bonds. For a flat linear interface, particles can only jump
transversely away the surface, $\nu$ equals the inverse of the lattice
coordination number, $q,$ and one may write $N_{\text{me,A}}\approx
4q^{-1}L_{\parallel}\exp(-2\bar{\Delta}/T)$ where $\bar{\Delta}$ is the mean
number of broken bonds per evaporation event. We multiplied here by 2 to take
into account evaporation of surface holes that travel within the stripe to
reach the (same) surface again. On the other hand, evaporation processes
induce changes $\Delta Y_{\text{cm}}=M^{-1}\delta y,$ where $\delta y$ is the
net particle (transverse) displacement, and $M\approx L_{\parallel}\times
\ell(t)$ for compact stripes. Therefore,
\begin{equation}
\mathcal{D}_{\ell}^{(\text{A})}\sim4q^{-1}\langle\delta y^{2}\rangle
\operatorname{e}^{-2\bar{\Delta}/T}L_{\parallel}^{-1}\ell^{-2}. \label{dA}%
\end{equation}

\noindent\ (B) A hole jumps one lattice spacing away within the stripe
interior. This induces $\Delta Y_{\text{cm}}=1/M$ or 0, depending on the jump
direction. One may write $N_{\text{me,B}}=2\nu\rho_{h}\left(  T\right)
L_{\parallel}\ell p_{h}\left(  T\right)  ,$ where $\rho_{h}$ is the density of
holes, $L_{\parallel}\ell$ is the \textit{volume} or total number of sites
within the liquid stripe, and $p_{h}$ is the jumping probability per unit
time. The factor 2 here comes from the fact that a hole modifies
$Y_{\text{cm}}$ when jumping to any of the two directions $\pm\vec{y}.$ At low
$T,$ $\rho_{h}$ is small; holes are then rather isolated from each other, so
that jumps do not modify the number of broken bonds, and $p_{h}\approx1.$ It
ensues
\begin{equation}
\mathcal{D}_{\ell}^{(\text{B})}\sim2q^{-1}\rho_{h}L_{\parallel}^{-1}\ell^{-1}.
\label{dB}%
\end{equation}

\noindent Note that a different dependence of (\ref{dA}) and (\ref{dB}) on
$\ell$ is a consequence of the fact that the rates $N_{\text{me,A}}$ and
$N_{\text{me,B}}$ involve processes consisting of evaporation on the line and
difussion on the bulk, respectively.

For $\rho=%
\frac12
,$ one has on the average stripes of width $\ell$ that are separated a
distance $\ell$ from each other. Therefore, a given stripe takes a mean time
$\tau_{\ell}=\ell^{2}/\mathcal{D}_{\ell}$ to find (and thus to coalesce with)
another one, and this causes its width to increase by $\Delta\ell=\ell.$
Consequently, $\operatorname{d}\ell/\operatorname{d}t\sim\Delta\ell
\ \tau_{\ell}^{-1}=\mathcal{D}_{\ell}\ell^{-1}.$ Together with (\ref{dA}) and
(\ref{dB}), respectively, this implies that mechanism A is characterized by a
power law $\ell\sim t^{1/4},$ and that mechanism B is to be associated with
$\ell\sim t^{1/3}.$ Furthermore, assuming that pattern growth in the DLG is
the result of competition between the two mechanisms, and that they are
independent of each other, $\mathcal{D}_{\ell}=\mathcal{D}_{\ell}^{(\text{A}%
)}+\mathcal{D}_{\ell}^{(\text{B})},$ it follows that
\begin{equation}
\frac{\operatorname{d}\ell}{\operatorname{d}t}\sim\frac{1}{L_{\parallel}%
}\left(  \frac{\alpha_{\text{A}}}{\ell^{3}}+\frac{\alpha_{\text{B}}}{\ell^{2}%
}\right)  , \label{kineq}%
\end{equation}
where $\alpha_{\text{A}}=4\nu\langle\delta y^{2}\rangle\operatorname{e}%
^{-2\bar{\Delta}/T}\ $and $\alpha_{\text{B}}=2\nu\rho_{h}.$ This is our
general result for the DLG as far as the field $E$ is large, e.g., infinite,
and the temperature $T$ is low enough so that the interfaces, and mechanisms A
and B, are sufficiently simple as assumed. This is to be compared with the
Lifshitz-Slyozov-Wagner behavior $\operatorname{d}\ell/\operatorname{d}%
t\sim\ell^{-2}$ which assumes spatial isotropy and diffusion directly governed
by surface tension. Formally, (\ref{kineq}) is similar to an equation obtained
before by assuming isotropic conditions; see \S 1.\cite{huse}

The consequences of (\ref{dA})--(\ref{kineq}) are as follows. Both (\ref{dA})
and (\ref{dB}) imply independently that
\begin{equation}
\ell\sim\left(  \varphi\theta\right)  ^{1/\varphi}\left(  t/L_{\parallel
}\right)  ^{1/\varphi}. \label{t1314}%
\end{equation}
The difference is that $\theta=\alpha_{A}$ and $\varphi=4$ from (\ref{dA})
while one obtains $\theta=\alpha_{B}$ and $\varphi=3$ from (\ref{dB}). On the
other hand, for sufficiently late times, $\ell$ becomes large and equation
(\ref{kineq}) simply solves into
\begin{equation}
\ell(t)\sim\alpha t^{1/3}+\zeta, \label{t13}
\end{equation}
where $\alpha^{3}=3\alpha_{\text{B}}L_{\parallel}^{-1}$ and $\zeta
=\alpha_{\text{A}}/2\alpha_{\text{B}}.$ That is, the prediction is that hole
diffusion within the stripe (mechanism B) will be dominant at late times. A
different hypothesis, based on the stripe evaporation picture, was shown in \cite{mukamel} 
to imply $\ell\sim\left(t/L_{\parallel}\right)  ^{1/3}$. 
This coincidence is not surprising since, as argued above, the coalescence of two 
particles stripes implies the evaporation of the intermediate hole stripe and due to the 
particle/hole simmetry in our system, both mechanisms (stripe diffusion/coalescence and stripe 
evaporation) correspond to the same physical process yielding the same behavior. 
In order to uncover the close analogy between the two pictures, one may notice that,
to evaporate a particle stripe, many of its particles must cross the 
surrounding hole stripes and stick on the neighboring particle stripes (this
is so since the particle density in the gas phase remains almost constant).
This particle migration process through the surrounding hole stripes is
in fact what we have called 'hole diffusion within the stripe' in the presence of
particle/hole simmetry. Hence the fundamental mechanism involved in a
stripe evaporation is the diffusion of its constituents through the neighboring
stripes. This observation is a key one to understand the relation between 
stripe's evaporation and hole (particle) diffusion.

The effect of mechanism A ---surface evaporation and subsequent
condensation--- on growth is more subtle. In fact, our theory predicts a
crossover from the $t^{1/4}$ regime to the $t^{1/3}$ regime as time is
increased. That is, the two mechanisms will have a comparable influence at
$t\sim\tau_{\text{cross}}$ with
\begin{equation}
\tau_{\text{cross}}=\frac{\left(  4\alpha_{\text{A}}\right)  ^{3}}{\left(
3\alpha_{\text{B}}\right)  ^{4}}L_{\parallel}. \label{time1}%
\end{equation}
For times $t<\tau_{\text{cross}}$, mechanism A is dominant and the $t^{1/4}$ behavior is 
expected, while mechanism B is dominant for $t>\tau_{\text{cross}}$ and the asymptotic
$t^{1/3}$ growth law is then observed.
The crossover time $\tau_{\text{cross}}$ is a macroscopic, observable time. Further, we may 
define the time $\tau_{\text{ss}}$ at which a single stripe is reached by the condition that
$\ell\left(  t\right)  \approx%
\frac12
L_{\perp}.$ One obtains%
\begin{eqnarray}
\tau_{\text{ss}} & = & \frac{L_{\parallel}}{\alpha_{\text{B}}} \{
\frac{L_{\perp}^{3}}{24}-\frac{\zeta L_{\perp}^{2}}{4}+2\zeta^{2}L_{\perp
} \nonumber \\ 
& & -8\zeta^{3} [  \ln\frac{\alpha_{\text{B}} (  2\zeta+ \frac12
L_{\perp})  }{L_{\parallel}}-\ln\frac{2\zeta\alpha_{\text{B}}
}{L_{\parallel}}] \}
\label{time2}
\end{eqnarray}
Hence our system is characterized by two different time scales, namely, $\tau_{\text{cross}}$
and $\tau_{\text{ss}}$. They depend on system size in a different way. 
For large systems one generally obtains $\tau_{\text{ss}} \gg \tau_{\text{cross}}$, so that
the system converges,
after a short, perhaps unobservable transient time, to the relevant $t^{1/3}$ 
behavior. However, there are small systems for which $\tau_{\text{ss}}<\tau_{\text{cross}}$. 
These systems will reach the stationary state (a single 
stripe) before having time to enter into the asymptotic $t^{1/3}$ regime. For
these small systems, the only relevant behavior is the $t^{1/4}$ one.
Therefore, there is  a {\it size} crossover between $t^{1/4}$ asymptotic 
behavior for small systems and 
$t^{1/3}$ asymptotic behavior for large ones. The condition 
$\tau_{\text{cross}}(T,L_{\parallel}) = \tau_{\text{ss}}(T,L_{\parallel},L_{\perp})$ 
defines the crossover size.

Consider now the parameter $\gamma\equiv\tau_{\text{cross}}\left(  T,L_{\parallel
}\right)  /\tau_{\text{ss}}\left(  T,L_{\perp},L_{\parallel}\right)  .$ It
follows that the $t^{1/3}$ behavior is dominant for $\gamma\ll1.$ However, one
also has that $\gamma\left(  T,L_{\perp},L_{\parallel}\right)  \rightarrow0$
for finite $T$ in the thermodynamic limit ($L_{\perp},L_{\parallel}%
\rightarrow\infty,\ L_{\perp}/L_{\parallel}=\operatorname{const}.).$
Consequently, the $t^{1/3}$ growth law is the general one, namely, the only
one we should expect to observe in a macroscopic system. Corrections to this
should only occur at early times in small systems. This is fully confirmed below.

One may also define a longitudinal length,\cite{prl01,mukamel} say
$\ell_{\parallel}\sim t^{a_{\parallel}},$ where one expects $a_{\parallel
}>1/3$ (given that the growth is more rapid longitudinally than transversely).
This length is only relevant during the initial regime, until stripes become
well-defined, all of them extending the whole length $L_{\parallel}$. This
condition may be taken as defining the onset of the multi-stripe state, which
may be characterized by $\ell_{\parallel}\left(  \tau_{\text{ms}}\right)
=L_{\parallel},$ from where it follows that $\tau_{\text{ms}}\sim
L_{\parallel}^{1/a_{\parallel}}.$ Interesting enough, this is on the
macroscopic time scale, as for both $\tau_{\text{cross}}$ and $\tau
_{\text{ss}}$ (more precisely, $\tau_{\text{cross}}\sim L_{\parallel}$ and
$\tau_{\text{ss}}\sim L_{\parallel}L_{\perp}^{3}).$ The fact that all these
relevant times are on the macroscopic, observable time scale confirms that, as
argued above, the single-stripe (and not the multi-stripe) state is the only
stable one in general. It is also to be remarked that, once the multi-stripe
state sets in, the only relevant length is the transverse one, $\ell.$ Of
course, this is compatible with the possible existence of two correlation
lengths describing thermal fluctuations at criticality. 
\begin{figure}
\centerline{
\psfig{file=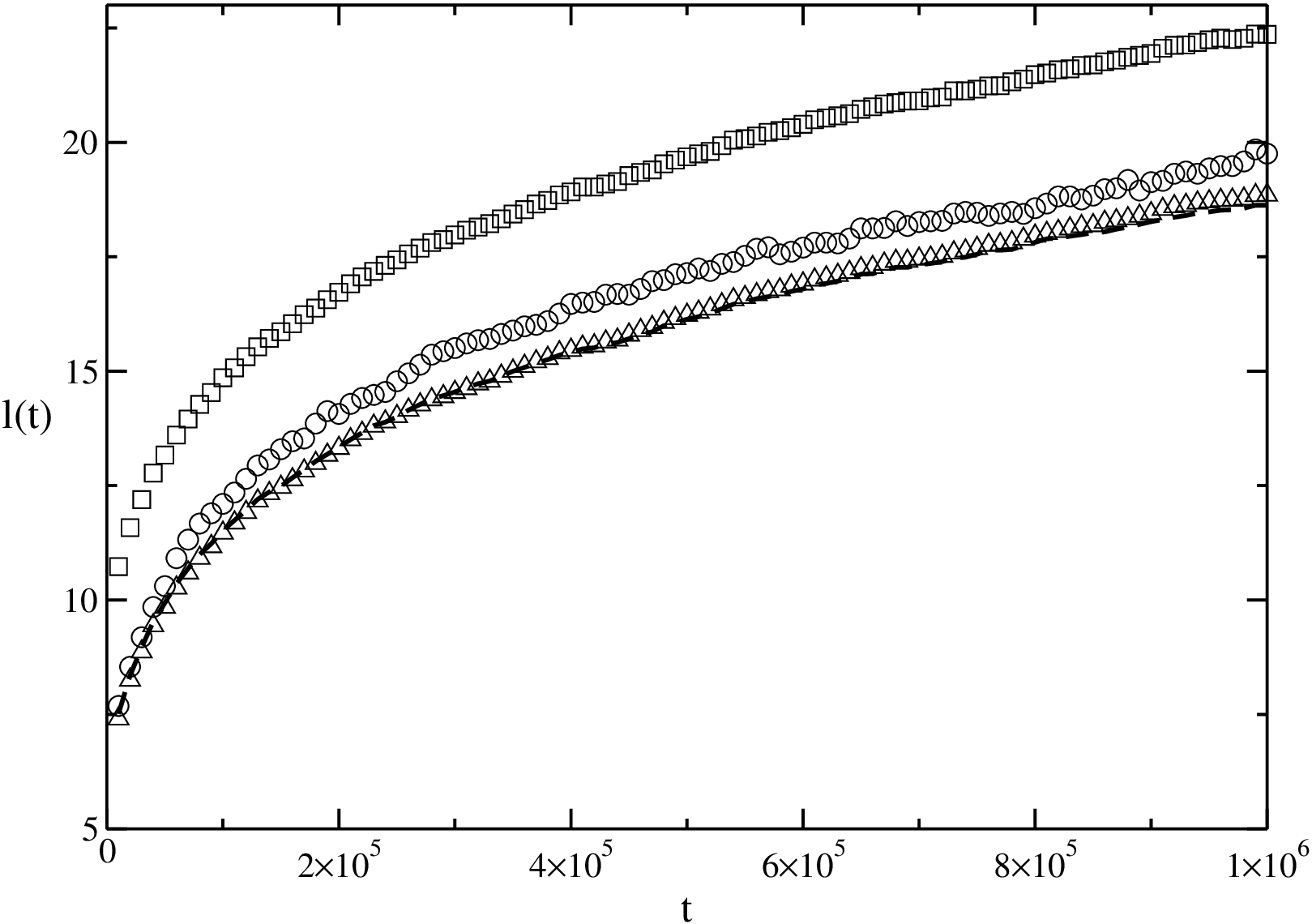,width=8.0cm,angle=0}}
\caption{\small Time evolution of the relevant length, $\ell\left(
t\right)$, as obtained by different methods, namely, from the
number $N_{s}$ of stripes (dashed line), from the maximum width,
$\ell_{\text{{\small max}}}$ $\left(  \Box\right)$ from the
mass, $\ell_{M}\left(  {\large \bigtriangleup}\right)$, and from
the peak of the structure function, $\ell_{S}\equiv2\pi/k_{\perp,\text{max}
}\left(  \bigcirc\right)$; these quantities are defined in the main
text. The graphs here correspond to an average over 600 independent runs for
the 128x128 lattice.\medskip}
\end{figure}

In order to test our predictions, several measures of the relevant length in
computer simulations were monitored, namely:

\begin{itemize}
\item the maximum width of the stripe, $\ell_{\text{max}},$ averaged over all
stripes in the configuration. This maximun width is defined as the distance in the
direction perpendicular to the field between the leftmost and the rightmost particles 
within the stripe;

\item $\ell_{M}\equiv M/L_{\parallel},$ where $M=M\left(  t\right)  $ is the
\textit{mass,} or number of particles belonging to the stripe, averaged over
all stripes in the configuration. This mass width is defined as the width of a 
perfectly dense stripe with $M$ particles.

\item $\ell_{s}\equiv L_{\perp}/2N_{s},$ where $N_{s}$ is the number of
stripes in the configuration.
\end{itemize}

\noindent After averaging over many independent evolutions, all these
quantities happen to behave similarly with time. Further measures of the
relevant length that we define in the next section behave in the same way. We
shall refer to this common behavior, which is illustrated in fig.3, as
$\ell\left(  t\right)  .$ (It is noticeable that, before showing a common
behavior, Fig.3 reveals some significant differences between our measures of
$\ell\left(  t\right)  $ at early times. This confirms the more difficult
description ---not attempted here--- which is required by the initial regime.) 
\begin{figure}
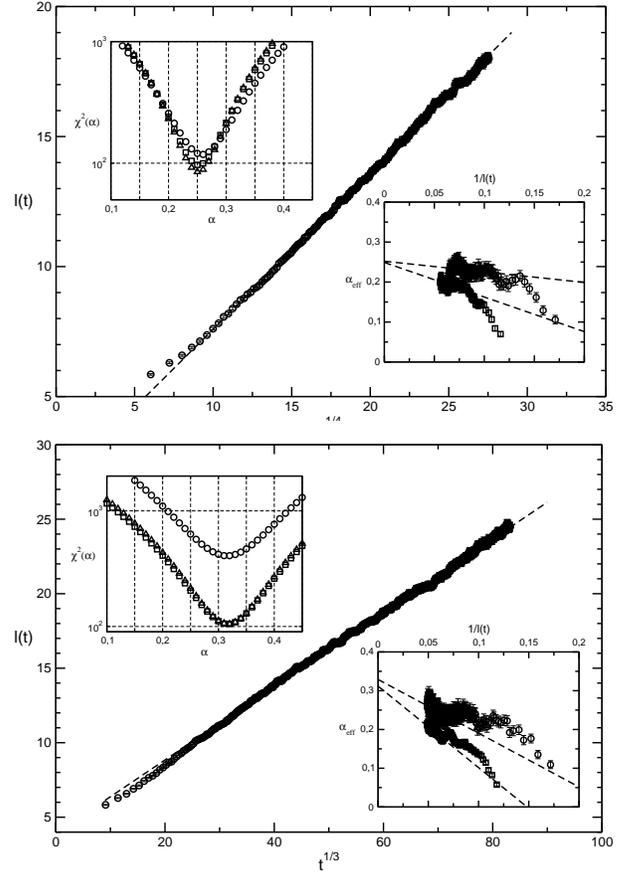

\centerline{
\psfig{file=PRBfigure4a-new.eps,width=8.0cm,angle=0}}
\centerline{
\psfig{file=PRBfigure4b-new.eps,width=8.0cm,angle=0}}
\caption{\small (a) The main graph shows $\ell\left(  t\right)  =\ell
_{S}\left(  t\right)$ versus $t^{a_{\perp}}$ for
$a_{\perp}=1/4$ in the case of the \textquotedblleft
small\textquotedblright\ 64x64 lattice. A similar behavior is obtained for any
of the studied measures of $\ell$ (see the main text for
definitions), which are represented in the insets by different symbols,
namely, $\ell_{\text{{\small max}}}$ (empty squares), $\ell_{S}$
(empty circles), and $\ell_{M}$ (triangles).
The upper inset shows the chi square function for varying $a_{\perp}%
$ as obtained from a series of fits; a well-defined minimum is
exhibited indicating that $a_{\perp}\simeq1/4$ in this case. The
lower inset shows the \textit{effective exponent,} $\text{d}\log
_{2}\ell\ /\ \text{d}\log_{2}t,$ as a function of 
$1/\ell\left(  t\right)  $; this extrapolates to the same value of
$a_{\perp}$. (b) Same as (a) but demonstrating that $a_{\perp}=1/3$
for the \textquotedblleft large\textquotedblright\ $L_{\perp}\times L_{\parallel} = 256x64$ 
lattice (one obtains a similar result for larger $L_{\parallel}$).medskip}
\end{figure}

In fig.4 we illustrate our analysis and main results concerning the (late)
time evolution of $\ell\left(  t\right)  .$ The predictions above are
confirmed and, in particular, \textquotedblleft small\textquotedblright%
\ lattices ---fig.4(a)--- happen to behave differently than \textquotedblleft
large\textquotedblright\ lattices ---fig.4(b). In both cases we plotted
$\ell\left(  t\right)  $ versus $t^{a_{\perp}}$ for varying $a_{\perp},$
looking for the best linear fit $\ell(t)=\alpha t^{a_{\perp}}+\zeta,$
excluding the initial time regime. The upper insets in the figures show the
chi square function associated to each fit, namely,
\begin{equation}
\chi^{2}\left(  a_{\perp}\right)  =\sum_{i=1}^{\eta}\frac{\left[  \ell
(t_{i})-\left(  \alpha t_{i}^{a_{\perp}}+\zeta\right)  \right]  ^{2}}{\alpha
t_{i}^{a_{\perp}}+\zeta},
\end{equation}
for a least-squares fit to $\eta$ data points using parameters $a_{\perp},$
$\alpha$ and $\zeta$. The main graphs confirm the existence of a common
behavior for all the monitored measures of $\ell\left(  t\right)  $ (indicated
by different symbols). These graphs also demonstrate that $\ell(t)=\alpha
t^{a_{\perp}}+\zeta,$ with small $\zeta,$ during the whole time regime of
consideration. On the other hand, the upper insets indicate that $a_{\perp}$
is very close to $\frac{1}{4}$ for \textquotedblleft small\textquotedblright%
\ systems (in fact, for $L_{\perp}\leq128$) while $a_{\perp}\simeq\frac{1}{3}$
as the system becomes larger, say $L_{\perp}\geq256$ that corresponds to a
\textquotedblleft large\textquotedblright\ lattice according to familiar MC
standards. As an alternative method to analyze $\ell\left(  t\right)  ,$ one
may evaluate
\begin{equation}
\overline{a}\left(  t\right)  \equiv\frac{\operatorname{d}\log_{n}\ell\left(
t\right)  }{\operatorname{d}\log_{n}t}.
\end{equation}
Our prediction is that $\overline{a}\left(  t\right)  =a_{\perp}-\zeta
a_{\perp}/\ell\left(  t\right)  ,$ i.e., this should provide the exponent
$a_{\perp}$ by extrapolating to large $\ell\left(  t\right)  $ (late time).
The insets at the bottom of each fig.4 show the results for $n=2.$ They are in
agreement with the other method, and again confirm our predictions.

As indicated above, the size crossover between the $t^{1/4}$ and $t^{1/3}$ asymptotic regimes
is expected for a system size $(L_{\parallel},L_{\perp})$ such that 
$\tau_{\text{cross}}\left(  T,L_{\parallel}\right)
=\tau_{\text{ss}}\left(  T,L_{\perp},L_{\parallel}\right)  .$ 
In order to make
this condition explicit, we need to estimate the amplitudes $\alpha_{\text{A}%
}\ $and $\alpha_{\text{B}}$ in (\ref{kineq}); see equations (\ref{time1}) and
(\ref{time2}). These amplitudes, which state the relative importance of
surface evaporation/condensation versus bulk hole-diffusion, are given
respectively by $\alpha_{\text{A}}=4q^{-1}\langle\delta y^{2}\rangle
\operatorname{e}^{-2\bar{\Delta}/T}\ $and $\alpha_{\text{B}}=2q^{-1}\rho_{h}.$
We note that, for a sufficiently flat interface (i.e., one that involves
microscopic --but not macroscopic-- roughness), $\langle\delta y^{2}%
\rangle\sim\mathcal{O}\left(  1\right)  $ and $\bar{\Delta}\simeq5.$ On the
other hand, the excess energy associated with an isolated hole is 16, so that
$\rho_{h}\sim\exp\left(  -16/T\right)  $ is a rough estimate of the hole
density. As depicted in fig.5, it follows numerically, in full agreement with
our observations, that $a_{\perp}=\frac{1}{4}$ is to be observed only at early
times, earlier for larger systems; to be more specific, the crossover for
$L_{\parallel}=64,$ for instance, is predicted for $L_{\perp}\sim140,$ which
confirms the above; see also figures 4.
\begin{figure}
\centerline{
\psfig{file=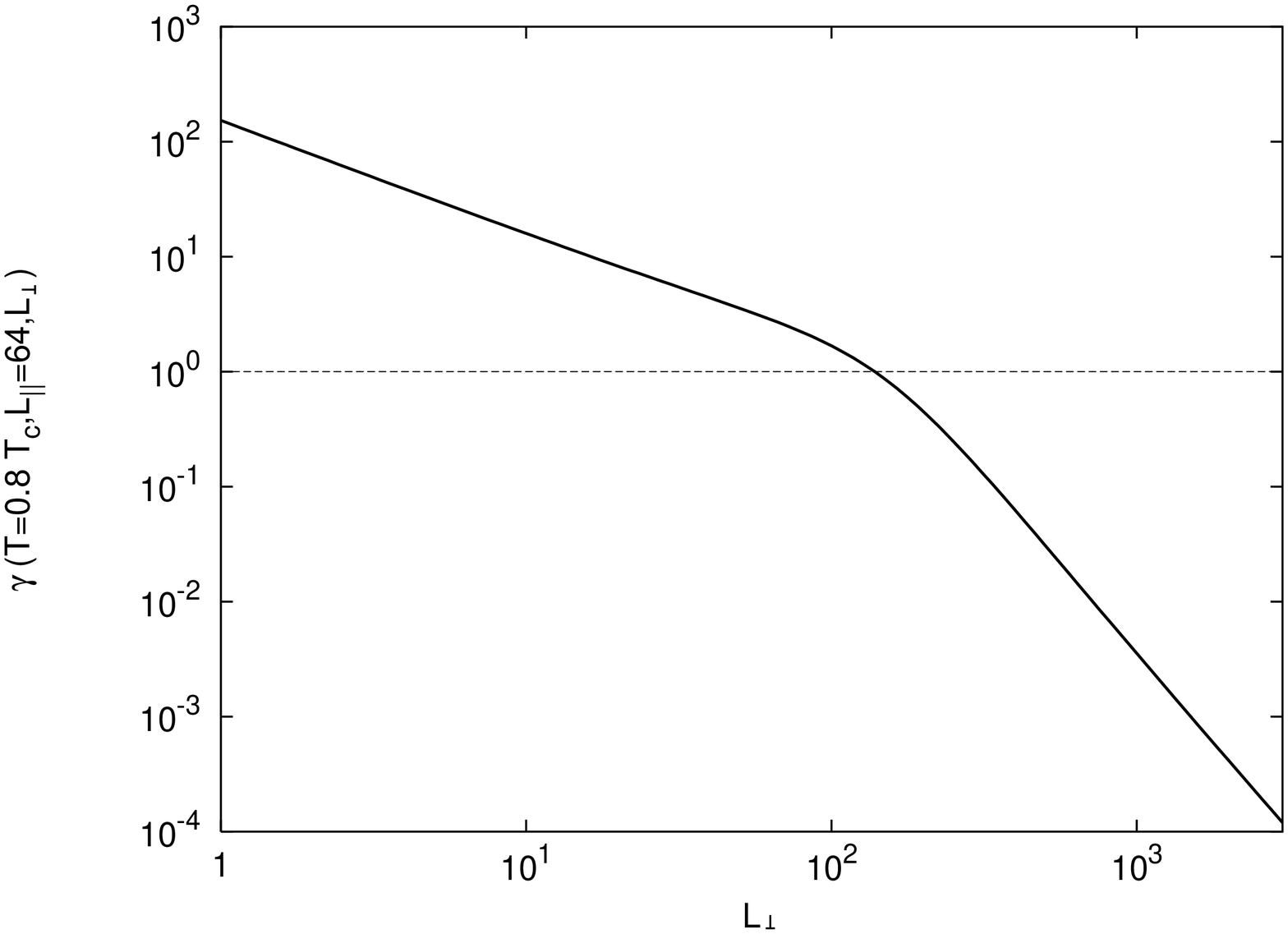,width=9.0cm}}
\caption{\small The parameter $\gamma=\tau_{\text{{\small cross}}
}(T,L_{\parallel})/\tau_{\text{{\small ss}}}(T,L_{\perp},L_{\parallel}),$
with the characteristic times $\tau_{\text{cross}}$ and
$\tau_{\text{ss}}$ defined in the main text, as a function of
$L_{\perp}$ for $L_{\parallel}=64.$ This confirms our
distinction between \textquotedblleft small\textquotedblright\ and
\textquotedblleft large\textquotedblright\ lattices, as explained in the main
text.\medskip}
\end{figure}

This behavior may be understood on simple grounds. The surface/volume ratio is
large initially and, consequently, mechanism A (based on surface events) is
then dominant. This is more dramatic the smaller the system is. That is, the
surface is negligible for macroscopic systems, in general, and, as illustrated
in fig.2, even if the surface is relevant at very early times, its ratio to
the volume will monotonically decrease with time. This causes hole diffusion
in the bulk (mechanism B) to become dominant, more rapidly for larger systems,
as the liquid phase is trying to exhibit only two surfaces.
On the other hand, ref. \cite{mukamel} 
studies the stripe coarsening process in the infinitely 
driven lattice gas. Pure $t^{1/3}$ behavior is reported assuming 
the stripe evaporation mechanism.
This result is perfectly compatible with our results, given that the systems in ref. 
\cite{mukamel} correspond to very large values of $L_{\perp}$ ($800$ and $960$) and small 
values of $L_{\parallel}$ ($8$, $16$ and $32$). For these shapes our theory also predicts
the (simple) $t^{1/3}$ asymptotic behavior.

\section{Correlations and the structure factor}

Consider now the Fourier transform of the pair correlation function
$C(x,y;t)=\left\langle n_{0,0}\left(  t\right)  \ n_{x,y}\left(  t\right)
\right\rangle ,$ where $n_{x,y}$ stands for the occupation variable at lattice
site $\vec{r}=\left(  x,y\right)  .$ This is the so-called structure factor,
$S(\vec{k},t),$ where $\vec{k}=(k_{\Vert},k_{\perp}).$ Given that the
$k_{\parallel}$ dependence is only relevant at early times, before the
multi-stripe state sets in, i.e., for $t<\tau_{\text{ms}}$, we shall set
$k_{\parallel}=0$ in the following. That is, our interest here is on%
\begin{equation}
S\left(  k_{\perp};t\right)  =\frac{1}{L_{\parallel}L_{\perp}}\left\vert
\sum_{x,y}n_{x,y}\left(  t\right)  \exp\left[  \text{i}k_{\perp}y\right]
\right\vert ^{2}.
\end{equation}
As illustrated in fig.6, this function develops a peak at $k_{\perp
}=k_{\text{max}}\left(  t\right)  $ immediately after quenching. The peak then
monotonically shifts towards smaller wave numbers with increasing $t;$ in
fact, one expects $k_{\perp}\rightarrow0$ as $t\rightarrow\infty$ in a
macroscopic system. The wave length $\ell_{S}\equiv2\pi/k_{\text{max}}$ turns
out to be an excellent characterization of the relevant order, namely, it
measures both the stripe width and the stripe separation during phase
segregation. In particular, we confirm that $\ell_{S}\left(  t\right)  $ has
the common behavior discussed above for length $\ell\left(  t\right)  ;$ see
figs. 3 and 4.

The fact that the DLG shows a unique \textit{time-dependent} relevant length,
$\ell_{\perp}=\ell\left(  t\right)  ,$ has some important consequences. For
example, extrapolating from the equilibrium case (see \S 1),\cite{prl00} one
should probably expect dynamical scaling, i.e.
\begin{equation}
S\left(  k_{\perp};t\right)  \propto\ell\left(  t\right)  F\left[  k_{\perp
}\ell\left(  t\right)  \right]
\end{equation}
for the anisotropic DLG in two dimensions. This is indeed observed to hold
during most of the relaxation and, in particular, during all the segregation
process after formation of well-defined stripes. This is illustrated in fig.7
depicting the scaling function $F.$ A time-dependent mean-field model of a
binary mixture in shear flow has recently been demonstrated to exhibit a
similar property, though involving two lengths both behaving differently from
$\ell\left(  t\right)  $ above.\cite{corberi} 
\begin{figure}
\centerline{
\psfig{file=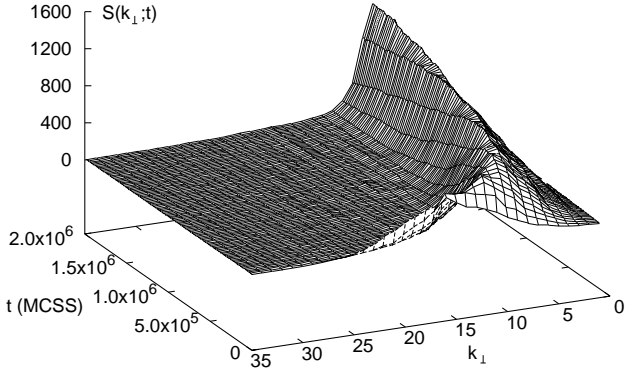,width=10.0cm}}
\caption{\small Time development of the structure factor $S(k_{\perp};t)$, as defined in the
main text, for a \textquotedblleft large\textquotedblright\ lattice 
$L_{\perp}\times L_{\parallel}= 256\times 256$ during
early and intermediate phase segregation. A peak grows with time as it shifts
towards the small values of $k_{\perp}.\medskip$}
\end{figure}

The structure factor may be obtained by scattering, which makes it an
important tool in many studies. Analyzing further the details of functions
$S\left(  k_{\perp};t\right)  $ and $F\left(  \varkappa\right)  $ or,
alternatively, the universal function $\Phi\left(  \varkappa\right)  \equiv
S/\ell L$, as observed in computer simulations is therefore of great interest
(the extra $L$ factor in the definition of $\Phi\left(  \varkappa\right)$ is
our finite size scaling ansatz).
Experimental studies often refer to the mean `radius of gyration' of the
grains as the slope of the straight portion in a plot of $\ln\left[  S\left(
k,t\right)  \right]  $ \textit{versus} $k^{2}.$\cite{guin} We checked the
validity under anisotropic conditions of this concept, which is in fact quite
useful in equilibrium even outside the domain of validity of its
approximations.\cite{prl00} We confirm that $S\left(  k_{\perp};t\right)  $
exhibits the Guinier Gaussian peak, namely,
\begin{equation}
\Phi\left(  \varkappa\right)  \sim\exp\left[  -\text{const.}\left(
\varkappa-\varkappa_{\text{max}}\right)  ^{2}\right]
\end{equation}
around the maximum $\varkappa_{\text{max}}.$ More intriguing is the behavior
of $\Phi\left(  \varkappa\right)  $ before the peak, $\varkappa<\varkappa
_{\text{max}}.$ Fig.7 indicates that scaling does not hold in this region even
at the end of our (otherwise long enough) simulations. This is so because
$\Phi\left(  \varkappa\right)  $ goes as $\rho^{2}L/\ell\left(  t\right)  $ at
$k_{\perp}=0$, 
and thus depends on time for very small values of $\kappa$, breaking the
scaling observed for larger values of $\kappa$.
However, a detailed study of data reveals that the scaling
function near the origin tends with time towards a common envelope
$\Phi\left(  \varkappa\right)  \sim\varkappa^{1+1/3}$ for $\varkappa
_{0}<\varkappa<\varkappa_{\text{max}};$ we do not have a simple explanation of
this. In any case, this behavior breaks down close to the origin,
$\varkappa\lesssim\varkappa_{0},$ where $\Phi\left(  \varkappa\right)
\rightarrow0$ as $\varkappa\rightarrow0$ and $t\rightarrow\infty$ for the
infinite system.
\begin{figure}
\centerline{
\psfig{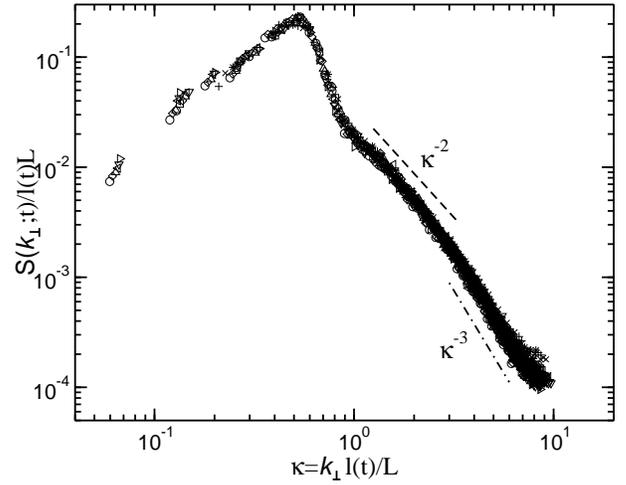}}
\caption{\small The scaling with both time and size of the structure
function to show that $\Phi(\kappa)\equiv S\left(  k_{\perp};t\right)
/\ell L,$ with $\kappa=k_{\perp}\ell L^{-1}$, is
well-defined and universal, i.e., the same at any time (excluding some early
evolution) and for any square lattice of side $L.$ This plot
includes all data for $t\geq$ 10$^{4}$ MC steps and 64x64,
128x128, and 256x256 lattices. The broken lines illustrate the different kinds
of behavior of $\Phi\left(  \kappa\right)$ that are discussed in
the main text.$\medskip$}
\end{figure}

The behavior after the peak, $\varkappa>\varkappa_{\text{max}},$ may be
predicted on simple grounds. The (sphericalized) structure factor for
(equilibrium) isotropic binary mixtures is known to satisfy the Porod's law,
$S\sim k^{-(d+1)}$ at large enough $k,$ where $d$ is the system
dimension,\cite{prl00} i.e., $S\sim k^{-3}$ in two dimensions. The main
contribution to the large-$k$ tail comes from the short-distance behavior of
$C(x,y;t).$ That is, the Porod's region for the DLG may be taken to correspond
to $\lambda_{\bot}\ll k_{\bot}^{-1}\ll\ell\left(  t\right)  ,$ where
$\lambda_{\bot}$ stands for a (transverse) thermal length that characterizes
the smallest, thermal fluctuations. Let two points, $\vec{r}_{0}$ and $\vec
{r}_{0}+\vec{r},$ $\vec{r}=(x,y).$ For any $x$ such that $\lambda_{\perp}\ll
x\ll\ell(t),$ one roughly has that the product $n_{\vec{r}_{0}}\left(
t\right)  \ n_{\vec{r}}\left(  t\right)  $ equals $+1$ if the two points are
on the stripe, and $0$ otherwise, i.e., if either an interface exists between
them or else the two points belong to the gas between stripes. Since $x\ll
\ell(t)$, the probability that $\vec{r}$ crosses more than one interface is
negligible. For a half-filled system, the probability that $\vec{r}_{0}$ lies
at a particle stripe is $%
\frac12
,$ and the probability that both $\vec{r}_{0}$ and $\vec{r}_{0}+\vec{r}$
belong to the same stripe is roughly $%
\frac12
\left(  \ell\left(  t\right)  -x\right)  /\ell\left(  t\right)  .$ Hence,
\begin{equation}
C\left(  x,y;t\right)  \simeq\frac{1}{2}\left(  1-\frac{x}{\ell\left(
t\right)  }\right)  ,\qquad{\small x\ll\ell\left(  t\right)  .}%
\end{equation}
By power counting, this implies the \textit{anisotropic Porod law} (in two
dimensions):%
\begin{equation}
S\left(  k_{\perp};t\right)  \sim\frac{1}{\ell\left(  t\right)  \ k_{\bot}%
^{2}},\qquad{\small \lambda_{\perp}\ll k_{\bot}^{-1}\ll\ell\left(  t\right)
.} \label{Porod}%
\end{equation}
Therefore, $\Phi\left(  \varkappa\right)  \sim\varkappa^{-2}L^{-1},$ which is
confirmed in fig.7. This is in contrast with the (isotropic) Porod's result.
The difference is a consequence of the fact that the DLG clusters are stripes
that percolate in the direction of the field, instead of the isotropic
clusters of the LG. The short-distance pair correlation function for the
latter is $C(\vec{r};t)\simeq%
\frac12
\left(  1-|r|/\ell\left(  t\right)  \right)  $, from which one has that
$\Phi\left(  \varkappa\right)  \sim\varkappa^{-3}.$ It follows that anisotropy
may easily be detected by looking at the tail of the structure factor.

The detailed analysis of $S\left(  k_{\perp};t\right)  $ also reveals that, as
$L_{\parallel}$ is increased in computer simulations, the anisotropic behavior
$\Phi\sim\varkappa^{-2}$ crosses over to $\Phi\sim\varkappa^{-3}$ for larger
$\varkappa$; see fig.7. We believe this reflects the existence of standard
thermal fluctuations. That is, very small clusters of particles occur in the
gas in the asymptotic regime whose typical size in the direction perpendicular
to the field is of order $\lambda_{\perp}$. These very-small asymptotic
clusters are rather isotropic, namely, they do not differ essentially from the
corresponding ones in equilibrium binary mixtures. More specifically, for
$x\sim\lambda_{\perp},$ one may approximate $C(\vec{r};t)\sim1-|r|/\lambda
_{\perp}(t),$ which implies the $\varkappa^{-3}$ power-law tail for large
$\varkappa.$ On the other hand, according to (\ref{t1314}), the mean stripe
width grows as $\ell(t)\sim(t/L_{\parallel})^{a}$ with $a=1/4$ or $a=1/3,$
depending on the value of $L_{\perp}$. Therefore, the number of stripes at
time $t$ is proportional to $L_{\perp}L_{\parallel}^{a}/t^{a}$ and, for a
given time, the number of stripes increases with $L_{\parallel}$ as
$L_{\parallel}^{a}$. We also know that, at a given time, the number of small,
fluctuating clusters is proportional to $L_{\parallel}$. Hence the relative
importance of small clusters due to thermal fluctuations as compared to
stripes is proportional to $L_{\parallel}.$ In fact, the $\varkappa^{-3}$ tail
is observed for large enough values of $L_{\parallel}$ but not for small lattices.

\section{A continuum description\label{Lang}}

The rigorous derivation of a general continuum analog of the driven lattice
gas is an open problem.\cite{md} Recent studies led to the following proposal
for a coarse-grained density, $\phi(\mathbf{r},t):$\cite{Rbien}
\begin{equation}
\partial_{t}\phi(\mathbf{r},t)=\tau_{\perp}\nabla_{\perp}^{2}\phi
-\nabla_{\perp}^{4}\phi+\frac{\lambda}{6}\nabla_{\perp}^2\phi^{3}+\tau
_{\parallel}\nabla_{\parallel}^{2}\phi+\nabla_{\perp}\xi(\mathbf{r}%
,t).\label{bien}%
\end{equation}
Here, the last term stands for a conserved Gaussian noise representing the
fast degrees of freedom, and $\tau_{\perp}$, $\tau_{\parallel}$ and $\lambda$
are model parameters. Compared to previous proposals,\cite{Rmal,zia} this
Langevin type of equation amounts to neglect a nonlinear current term,
$-\alpha\nabla_{\parallel}\phi^{2},$ that was believed to be essential
(\textit{relevant}) at criticality. However, one may show that, at least in
the limit $E\rightarrow\infty,$ the coefficient $\alpha$ cancels out (due in
this case to a subtle saturation effect).\cite{Rbien} In fact, recent scaling
analysis has unambiguously confirmed that a particle current is not relevant
and that equation (\ref{bien}) captures the correct critical behavior of the
DLG.\cite{prl01,albano} Consequently, an important question is now whether
(\ref{bien}) reproduces also the kinetic behavior of the DLG as described in
previous sections. We present in this paper a first confirmation that, as
compared with other approaches,\cite{zia} (\ref{bien}) is indeed a proper
continuum description of the DLG kinetic relaxation. A more complete study of
the kinetic consequences of this equation will be presented elsewhere.
\begin{figure}
\centerline{
\psfig{file=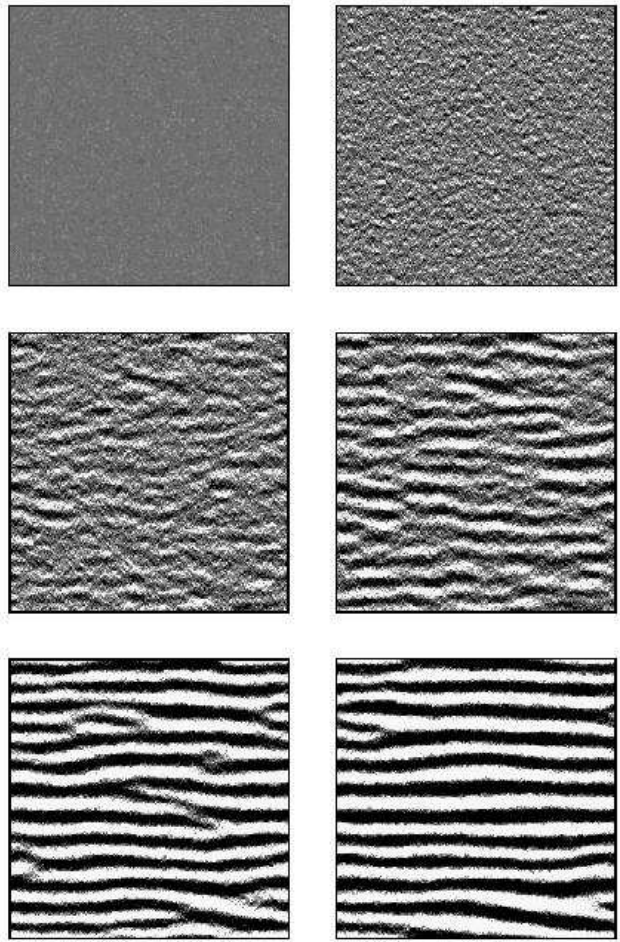,width=9.0cm,angle=0}}
\caption{\small Series of snapshots as obtained from equation (\ref{discrte}) 
for the $256\times256$ lattice with parameters as given in the main text.
Time (arbitrary units) is $t=0,10,100,200,500$, and $1000$ respectively, from
left to right and from top to bottom.$\medskip$}
\end{figure}

In order to numerically integrate (\ref{bien}), let us introduce the indexes
$i,j=1,\dots,N$ to represent, respectively, the two components of
$\mathbf{r\equiv}\left(  x_{\perp},x_{\parallel}\right)  .$ One thus makes a
trivial discretization of the space, and then of the time by Cauchy-Euler
method.\cite{CEmethod} The result is a set of $N^{2}-1$ coupled nonlinear
equations, namely,%
\begin{eqnarray}
\phi(i,j;t+\Delta t) &  = & \phi(i,j,t)\label{discrte}
+\Delta t [  \tau_{\perp}\widetilde{\nabla}_{\perp}^{2}\phi
-\widetilde{\nabla}_{\perp}^{4}\phi+\frac{\lambda}{6}\widetilde{\nabla}
_{\perp}^{2}\phi^{3} \nonumber \\
& & +\tau_{\parallel}\widetilde{\nabla}_{\parallel}^{2}%
\phi ]  +\sqrt{\Delta t}\widetilde{\nabla}_{\perp}\xi(i,j;t).
\end{eqnarray}

This equation is to be solved by the computer. With this aim, we may write
$\widetilde{\nabla}_{\perp}\xi(i,j;t)=\left[  \xi(i+1,j;t)-\xi
(i-1,j;t)\right]  /2\Delta x_{\perp}$ and $\phi(i,j;t)\equiv\phi(i\Delta
x_{\perp},j\Delta x_{\parallel};t)$ where $\Delta x_{\perp}=L_{\perp}/N$ and
$\Delta x_{\parallel}=L_{\parallel}/N.$ The maximum value of $\Delta x_{\perp
}$ is thus limited by the interface width. For fig.8, which concerns a
$256\times256$ lattice $(N=256)$ we ---rather arbitrarily--- used $\Delta
x_{\perp}=\Delta x_{\parallel}=1.7,$ and $\Delta t=0.05,$ which produce a
locally stable solution. The parameters $\tau_{\perp}$, $\tau_{\parallel}$ and
$\lambda$, are fixed on the basis of its physical meaning. The \textit{mass}
terms $\tau_{\parallel}$ and $\tau_{\perp}$ represent temperatures along the
longitudinal and transverse directions, respectively, relative to the critical
temperature, i.e., $\tau_{\perp}\sim(T_{\perp}-T_{C}^{\infty}).$ Given the
anisotropy of phase segregation, with longitudinal interfaces only,
$\tau_{\perp}<0$ and $\tau_{\parallel}>0.$ On the other hand, $\left\vert
\tau_{\perp}\right\vert $ should be small enough to allow for a relatively
fast evolution. Our choices for fig.8 are $\tau_{\perp}=-0.25$, $\tau
_{\parallel}=0.5$ and $\lambda=1$.

It is remarkable that, in spite of some apparent similarity, the
problem here differs from the one in the study of (standard) spinodal
decomposition by means of the isotropic ($E=0$) Cahn-Hilliard equation. In
equilibrium,\cite{CHeq} one usually assumes that the influence of noise on
growth, which is then assumed to be directly driven by surface tension, is
negligible far from criticality. The noise term in (\ref{discrte}) may be
expected to be important in a more general context, however. That is, as
described in \S \ref{growth}, the DLG develops striped patterns in which
surface tension smooths the interfaces but has no other dominant role on the
basic kinetic events. Consequently, neglecting the noise in (\ref{discrte})
would turn metastable any striped geometry after coarsening of strings, which
is not acceptable (see \S \ref{growth}).

Finally, it  is  interesting  to notice that  if a one-dimensional structure is assumed, and 
the gradient in the direction parallel to the field  in  Eq. \ref{bien} is  eliminated,
then  this equation reduces to the one-dimensional time-dependent Ginzburg-Landau model 
in \cite{kawakatsu}. 
There  it was found a $\ln(t)$ growth at zero temperature  and a crossover
from $\ln(t)$ to $t^{1/3}$ at finite temperatures.

\section{Conclusion}

This paper presents a theoretical description of spinodal decomposition in the
DLG, and compares it with new data from a kinetic Monte Carlo study. This is
also compared with the kinetic implications of a Langevin, continuum equation
that had previously been shown to capture correctly the critical behavior of
the DLG. The resulting picture from these three approaches, which is
summarized below, should probably hold for a class of higly-anisotropic phase
segregation phenomena. In fact, our results provide a method for analyzing
experiments that could be checked against laboratory realizations of the DLG,
i.e., the case of phase segregation under biased fields or other influences
such as electric fields, gravity and elastic stresses.

Immediately after a deep quench, there is an early regime in which anisotropic
grains develop. They tend to coarsen to form small strings that then combine
into well-defined thin stripes. Such nucleation and early coarsening (fig.1)
seem governed by surface tension at the string ends competing with other both
surface and bulk processes. This complicated situation typically extends less
than 10$^{3}$ MC steps in computer simulations, which corresponds to a very
short macroscopic time, so that it would be hardly observable in experiments.
As a matter of fact, most of the system relaxation proceeds by coarsening of
stripes until full segregation (fig.2). Surprisingly enough, this regime,
which has been studied for more than a decade now,\cite{aurora}-\cite{mukamel}
happens to be theoretically simpler than the corresponding one for the
isotropic case.\cite{rev1}-\cite{toral}

The evolution from many stripes to a single one mainly proceeds by competition
of two mechanisms: (A) evaporation of a particle from one stripe surface and
subsequent deposition at the same surface, and (B) diffusion of a hole within
the bulk of the stripe. The first one dominates initially (and lasts more for
smaller systems), when the surface/volume ratio is relatively large. Mechanism
A implies that the relevant length (as defined in fig.3) increases with time
according to $\ell\left(  t\right)  \sim t^{1/4}.$ The surface/volume ratio
decreases with time, however, and mechanism B soon becomes dominant. This
implies $\ell\left(  t\right)  \sim t^{1/3}$ which is the general prediction
for a macroscopic system (cf. figures 4 and 5).\cite{voter} This was obtained
before by assuming coarsening of two (liquid) stripes by evaporation of the
gas stripe placed between them;\cite{mukamel} see also \cite{aurora} Note that
the $t^{1/3}$ law is precisely the behavior which is acknowledged to be
dominant under isotropy, but this has a different origin in the equilibrium
case.\cite{rev4,rev5} Note also that surface tension determines evaporation
rates but has no other influence on mechanisms A and B.

The $t^{1/3}$ growth law, (\ref{t13}), is perfectly confirmed by the DLG data
(fig.4b). This indicates time-scale invariance. In fact, such invariance was
demonstrated for the isotropic case, in which the situation is somewhat more
involved (\S 1). The invariance property may be better analyzed by looking at
the structure factor transversely to the drive, $S\left(  k_{\perp},t\right)
$ (fig.6). This exhibits \textit{dynamic scaling}, i.e., it remains
self-similar during phase segregation.\cite{man} More specifically,
$\Phi(\varkappa)\equiv S\left(  k_{\perp};t\right)  /\ell L,$ with
$\varkappa=k_{\perp}\ell L^{-1}$, is universal, namely, the same at any
(sufficiently late) time $t$ and for any square lattice of side $L.$
Furthermore, the function $\Phi(\varkappa)$ has a well-defined shape. In
particular, it exhibits the Guinier Gaussian peak, and this is followed by the
\textit{anisotropic Porod} decay, $\Phi\left(  \varkappa\right)  \sim
\varkappa^{-2}$ and then by a \textit{thermal tail} $\Phi\left(
\varkappa\right)  \sim\varkappa^{-3}$ (fig. 7). Also noticeable is the fact
that the the parameter to scale along the $S$ axis is $J(t)=\ell$ and not
$\ell^{2}$ as under isotropy.

Our results in this paper have two main restrictions, both due to the great
computational effort required by this problem.\cite{md} Firstly, they follow
from an extensive analysis of only one phase-diagram point, namely, $\rho=%
\frac12
,$ $E=\infty,$ and $T=0.8T_{C}^{0}.$ However, our own observations (including
brief investigation of other points), together with an extrapolation of the
many results known for the isotropic case, strongly suggest that the picture
in this paper holds within a large domain around the center of the miscibility
gap.\cite{fuera} In fact, the scaled structure factor for isotropic systems
was shown to be almost independent of density and temperature, and even the
substance investigated, in a wide region below the coexistence line.\cite{LSWbis} 
Our consideration of only a two-dimensional system
does not seem a real restriction neither. That is, adding an extra
(transverse) dimension should not essentially modify the picture
here.\cite{LSWter}

It would be interesting to look next in the laboratory for both time-scale
invariance and $t^{1/3}$ growth under highly anisotropic conditions. In fact,
there are some evidences of such behavior in sheared fluids (\S 1), and one
may think of some more direct experimental realizations of the driven lattice
gas. In particular, coarsening striped patterns very similar to those observed
in our system are found in some intringuing experiments on granular binary mixtures 
under shaking.\cite{Mullin} We think that the mechanisms we propose in this paper should help
the understanding of such experimental results. In general,
we hope our observations will motivate both experiments and future more
complete theories.

\section*{Acknowledgments}

We acknowledge very useful discussions with Ra\'{u}l Toral and Miguel
\'{A}ngel Mu\~{n}oz, and support from MCYT, projects PB97-0842 and BFM2001-2841.
We also acknowledge some useful comments by the referees.


\begin{thebibliography}{99}
\bibitem {man}More properly, one should speak here of \textquotedblleft
self-affinity\textquotedblright, given the underlying anisotropy.\cite{mandel}
\bibitem {fuera}Extending some of our arguments to $\rho\neq\frac{1}{2}$ needs
some care but the whole picture should still be valid, at least not far from
$\rho=\frac{1}{2}.$
\bibitem {LSWter}Note that the Lifshitz-Slyozov-Wagner behavior is known to be
valid in some cases in both $d=2$ and $d=3$.\cite{rev5}
\bibitem{interfases} Some partial, unconclusive studies about interfaces in the Driven 
Lattice Gas\cite{interf} have reviewed in Ref. \cite{md}.
\end{thebibliography}
\end{document}